\documentclass[pra,twocolumn,showpacs,amsmath,superscriptaddress,amssymb]{revtex4}
\usepackage{graphicx}
\usepackage{dcolumn}
\usepackage{bm}
\usepackage{epsfig}
\usepackage{subfigure}

\begin{document}

\title{Effects of disorder on quantum fluctuations and superfluid density of a Bose-Einstein condensate in a
two-dimensional optical lattice}

\author{Ying Hu  }
\affiliation{Department of Physics, Centre for Nonlinear Studies,
and the Beijing-Hong Kong-Singapore Joint Centre for Nonlinear and
Complex Systems (Hong Kong), Hong Kong Baptist University, Kowloon
Tong, Hong Kong, China}

\author{Zhaoxin Liang  }
\affiliation{Department of Physics, Centre for Nonlinear Studies,
and the Beijing-Hong Kong-Singapore Joint Centre for Nonlinear and
Complex Systems (Hong Kong), Hong Kong Baptist University, Kowloon
Tong, Hong Kong, China}

\affiliation{Shenyang National Laboratory for Materials Science,
Institute of Metal Research, Chinese Academy of Sciences, Wenhua
Road 72, Shenyang 110016, China}

\author{Bambi Hu }
\affiliation{Department of Physics, Centre for Nonlinear Studies,
and the Beijing-Hong Kong-Singapore Joint Centre for Nonlinear and
Complex Systems (Hong Kong), Hong Kong Baptist University, Kowloon
Tong, Hong Kong, China} \affiliation{Department of Physics,
University of Houston, Houston, TX 77204-5005, USA}

\date{\today}
\begin{abstract}
We investigate a Bose-Einstein condensate trapped in a 2D optical
lattice in the presence of weak disorder within the framework of the
Bogoliubov theory. In particular, we analyze the combined effects of
disorder and an optical lattice on quantum fluctuations and
superfluid density of the BEC system. Accordingly, the analytical
expressions of the ground state energy and quantum depletion of the
system are obtained. Our results show that the lattice still induces
a characteristic 3D to 1D crossover in the behavior of quantum
fluctuations, despite the presence of weak disorder. Furthermore, we
use the linear response theory to calculate the normal fluid density
of the condensate induced by disorder. Our results in the 3D regime
show that the combined presence of disorder and lattice induce a
normal fluid density that asymptotically approaches $4/3$ of the
corresponding condensate depletion. Conditions for possible
experimental realization of our scenario are also proposed.
\end{abstract}

\maketitle

\section{introduction}

Ultracold atoms in an optical lattice have opened a new theoretical
and experimental window for investigating fundamental problems
related to condensed matter physics in a very versatile manner
\cite{Morschrev,Blochrev,Yukalov}. Bose-Einstein condensates (BEC)
trapped in an optical lattice allow for optimal control of the
system's parameters: the depth of an optical lattice can be
arbitrarily modified by changing the intensities of laser beams
\cite{Greiner}, whereas the interatomic interactions can be
controlled via the technology of Feshbach resonances
\cite{Tiesinga}. Having at hand the possibility to shape the optical
lattices and realize quasi-1D, quasi-2D and 3D BEC systems almost at
will, therefore, an important direction of investigation consists in
studying the properties of a BEC system along the dimensional
crossover.

Along this line, much work have been done \cite{Morschrev,Blochrev}.
In particular, Orso {\it et al.}\cite{Orso} theoretically showed
that a 2D lattice can induce a characteristic 3D to 1D crossover in
the behavior of quantum fluctuations. Interest in this dimensional
crossover needs to be renewed, however, following the remarkable
observation that disorder in quantum systems can have dramatic
effects on the quantum fluctuations and superfluid density of a BEC
\cite{Yukalovd, Kuhn, Gimperlein, Sanchez, Delande,
Pilati,Fontanesi,Dang,Falco}. For example, it has been observed that
even a tiny amount of disorder in the confining fields leads to a
fractioning of quasi-1D condensates in waveguide structures on atom
chips \cite{Wang}. Hence, a natural question that immediately arises
is how the disorder affects the 2D-lattice-induced dimensional
crossover predicted in Ref. \cite{Orso}.

In this paper, we launch a systematic investigation on a BEC in the
combined presence of 2D optical lattice and weak external
randomness. The present work is divided into two parts. In the first
part, we shed new light on the interplay between disorder and
interatomic interaction along the 2D-lattice-induced dimensional
crossover. Accordingly, the analytical expressions of the ground
state energy and quantum fluctuations are derived for a BEC in a
tight 2D optical lattice in the presence of disorder using the
Bogoliubov theory. Our results show that for a fixed atomic density,
the BEC at small values of the lattice depth is in an anisotropic 3D
regime. When the lattice depth increases, however, the BEC undergoes
a dimensional crossover from an anisotropic 3D regime to a 1D
regime. The effects of disorder on such dimensional crossover is
analyzed in detail. Our results for the ground state energy in the
asymptotic 1D regime without disorder is in agreement with the exact
Lieb-Linger solution expanded in the weak coupling regime
\cite{Lieb}. We argue, therefore, that our result generalizes the
exact ground state energy of Lieb-Linger model to that in the
presence of weak disorder. All of our results in case of vanishing
disorder are consistent with those obtained in Ref. \cite{Orso}.

In the second part of this work, we calculate the normal fluid
density of the condensate induced by disorder using the linear
response theory. Accordingly, the transverse current-current
response functions are calculated. These results in case of
vanishing lattice depth recover corresponding ones in Ref.
\cite{Huang}. In particular, our results in the 3D regime show that
the combined presence of disorder and lattice induce a normal fluid
density that asymptotically approaches $4/3$ of the corresponding
condensate depletion \cite{Huang}.

The above scenario for disorder can be realized in cold atomic
systems using several methods in a controlled way. They include
applying optical potentials created by laser speckles or
multi-chromatic lattices \cite{Damski,Roth}, introducing impurity
atoms in the sample \cite{Ospelkaus} and manipulating the collision
between atoms \cite{Gavish}. The controllability of disorder in a
BEC system makes the studies of disorder-induced effects extremely
fascinating \cite{Lye,White,Billy}.

The outline of this paper is as follows. In Sec. II, we introduce
the Hamiltonian for a BEC trapped in a 2D optical lattice in the
presence of weak disorder and then analyze this model using the
Bogoliubov approximation. In Sec. III, the analytical expressions
for the ground state energy and quantum depletion of such BEC system
are obtained. Especially, we focus on the analysis of the combined
effects of disorder and optical lattice on quantum fluctuations of
this BEC system . Sec. IV presents calculations of reduction of the
superfluid density due to disorder trapped in an optical lattice.
Finally in Sec. V, we summarize our results and propose experimental
conditions for realizing our scenario.

\section{Hamiltonian for a BEC in both presence of a 2D optical lattice and weak disorder}

The N-body Hamiltonian describing the Bose system has the form
\begin{eqnarray}\label{Ham}
H-\mu N=\!\int
d{\bf{r}}\hat{\Psi}^{\dagger}({\bf{r}})\Big[\!\!&-\!&\!\frac{\hbar^2\nabla^2}{2m}-\mu\!+\!V_{opt}({\bf
r})+V_{ran}({\bf
r})\nonumber\\
&+&\frac{g}{2}\hat{\Psi}^{\dagger}({\bf{r}})\hat{\Psi}({\bf{r}})\Big]\hat{\Psi}({\bf{r}}),
\end{eqnarray}
where $\hat{\Psi}({\bf{r}})$ is the field operator for bosons with
mass $m$, $\mu$ is the chemical potential, $N=\int d {\bf
r}\hat{\Psi}^{\dagger}({\bf{r}})\hat{\Psi}({\bf{r}})$ is the number
operator, and $g=4\pi \hbar^2 a/m$ is the coupling constant with $a$
being the 3D scattering length in the free space. In the Hamiltonian
(\ref{Ham}), $V_{ran}({\bf r})$ and $V_{opt}({\bf r})$ respectively
represent the external random potential and the 2D optical lattice.

The 2D optical potential $V_{opt}({\bf r})$ in the Hamiltonian
(\ref{Ham}) is given by
\begin{equation}
V_{opt}({\bf r})=sE_R\left[\sin^2(q_Bx)+\sin^2(q_By)\right],
\end{equation}
where $s$ is a dimensionless factor labeled by the intensity of the
laser beam and $E_R=\hbar^2q^2_{B}/2m$ is the recoil energy with
$\hbar q_B$ being the Bragg momentum. The lattice period is fixed by
$\pi/d$. Atoms are unconfined in the $z$-direction.

Disorder in the Hamiltonian (\ref{Ham}) is produced by the random
potential associated with quenched impurities \cite{Astra}
\begin{equation}\label{Ran}
V_{ran}({\bf r})=\sum_{i=1}^{N_{imp}}v\left(|{\bf r-r_{i}}|\right),
\end{equation}
with $v({\bf r})$ describing the two-body interaction between bosons
and impurities, ${\bf r}_i$ being the randomly distributed positions
of impurities and $N_{imp}$ counting the number of ${\bf r}_i$. To
obtain the concrete form of the pair potential $v({\bf r})$, we need
to investigate the scattering problem between a particle and a
quenched impurity whose mass is taken to be infinite in the presence
of 2D optical lattice. Here, we restrict ourself to the conditions
of a dilute BEC system in the presence of a very small concentration
of disorder. Thereby, the potential $v({\bf r})$ can be expressed by
a pseudo-potential $v({\bf r})=\tilde{g}_{imp}\delta({\bf r})$.
Here, the $\tilde{g}_{imp}$ is the effective coupling constant that
reads $\tilde{g}_{imp}=2\pi \hbar^2 \tilde{b}/m$, where $\tilde{b}$
represents the effective scattering length accounting for the
presence of a 2D optical lattice \cite{Wouters}.

In what follows, we will assume that the optical lattice is strong
enough to create many separated wells that give rise to an array of
condensates. Meanwhile, because of the quantum tunneling, the
overlap between the wave functions of two consecutive wells are
still sufficient to ensure full coherence even in the presence of
disorder. In such case, one is allowed to use the Bogoliubov theory
to study both equilibrium and dynamic behavior of the system at zero
temperature. In addition, we also suppose that the chemical
potential $\mu$ is small compared to the inter-band gap. By this
assumption, we restrict ourselves to the lowest band, where the
physics is governed by the ratio between the chemical potential
$\mu$ and the bandwidth $8t$, where $t$ is the tunneling rate
between neighboring wells. Generally speaking, for $\mu \ll 8t$, the
system retains an anisotropic 3D behavior, whereas for $\mu \gg 8t$,
the system undergoes a dimensional crossover to a 1D regime.

In the tight-binding approximation \cite{Kramer}, the lowest Bloch
band of the BEC system can be written in terms of Wannier functions
as $\phi_{k_x}(x)\phi_{k_y}(y)$, where $\phi_{k_x}(x)=\sum_l
e^{ilkx}w(x-ld)$ and
$w(x)=\exp[-u^2/2\sigma^2]/\pi^{1/4}\sigma^{1/2}$ with
$d/\sigma\simeq \pi s^{1/4}\exp(-1/4\sqrt{s})$. Expanding the field
operators by the expression $\hat{\Psi}({\bf x})=\sum_{{\bf
k}}\hat{a}_{{\bf k}}e^{-ik_zz}\phi_{{k_x}}\left(
x\right)\phi_{{k_y}}\left( y\right)$, the Hamiltonian (\ref{Ham})
takes the form
\begin{eqnarray}
H^{'}&=&H-\mu N\nonumber\\
&=&\sum_{{\bf k}}\left(\epsilon_{{\bf
k}}^{0}-\mu\right)\hat{a}^{\dagger}_{{\bf k}}\hat{a}_{{\bf
k}}+\frac{\tilde{g}}{2V}\sum_{\bf k,q,k^{'}}\hat{a}^{\dagger}_{{\bf
k+q}}\hat{a}^{\dagger}_{{\bf k^{'}-q}}\hat{a}_{{\bf
k^{'}}}\hat{a}_{{\bf k}}\nonumber\\
&+&\sum_{\bf{k,k^{'}}}\hat{a}_{{\bf k}}^\dag \hat{a}_{\bf k^{'}}
{V}_{{\bf k}-{\bf k^{'}}},\label{HS}
\end{eqnarray}
where $\epsilon_{{\bf
k}}^{0}=\hbar^2k^2_z/2m+2t[2-\cos(k_xd)-\cos(k_yd)]$ is the energy
dispersion of the noninteracting model, $V$ is the volume and
$\tilde{g}=4\pi\hbar^2\tilde{a}/m$ is an effective coupling constant
with $\tilde{a}=C^2a$ and $C=d\int^{d/2}_{-d/2}w^4(x)dx\simeq
d/\sqrt{2\pi}\sigma$. In Eq. (\ref{HS}), the $V_{{\bf k}}$ is the
Fourier transform of $V_{ran}({\bf r})$
\begin{equation}\label{Vk}
V_{{\bf k}}=\frac{1}{V}\int e^{i{\bf k}\cdot {\bf r}}V_{ran}({\bf
r})d{\bf r}.
\end{equation}
Here we choose that the randomness is uniformly distributed with
density $n_{imp}=N_{imp}/V$ and Gaussian correlated \cite{Astra}.
Thereby, the two basic statistical properties of disorder are the
average value $\langle V_0\rangle=\tilde{g}_{imp}n_{imp}$ and the
correlation function $\langle V_{\bf k} V_{-\bf
{k}}\rangle=\tilde{g}^2_{imp}n_{imp}/V$. Here, the notation $\langle
... \rangle$ stands for the ensemble average over all disorder
configurations \cite{Yukalovd}.

In the following, we focus on the situation where the number of
atoms in each tube is sufficiently large and refer to $n_0$  as the
condensate density. Under this assumption \cite{Kramera}, we can
neglect the Mott insulator phase transition which would occur only
for extremely large values of the lattice depth. Throughout this
paper, we consider the case when the strength of disorder is weak
enough and Bogoliubov approximation still holds \cite{Astra}. Hence,
by proceeding in a standard fashion of Bogoliubov theory, the
Hamiltonian (\ref{HS}) can be approximated as
\begin{widetext}
\begin{eqnarray} \label{H2nd_2}
H_{eff}-\mu N&=&V\left(-\mu n_0+n_0\tilde{V}_0+\frac{1}{2}\tilde
gn^2_0\right)+\frac{1}{2} \sum_{{\bf k}\neq 0}
\left(\varepsilon_{\bf k}^0-\mu+2\tilde{g}n_0
\right)\left(\hat{a}_{\bf k}^\dag \hat{a}_{\bf k}+\hat{a}_{-{\bf
k}}^\dag \hat{a}_{-{\bf k}}\right)\nonumber\\
&+&\frac{1}{2}\tilde{g}n_0 \sum_{{\bf k}\neq 0} \left(\hat{a}_{\bf
k}^\dag \hat{a}_{-{\bf k}}^\dag+\hat{a}_{{\bf k}}\hat{a}_{-{\bf
k}}\right)+\sqrt{n_0} \sum_{{\bf k}\neq 0}\left(\hat{a}_{\bf k}^\dag
V_{-{\bf {\bf k}}}+\hat{a}_{{\bf k}}V_{{\bf k}}\right)+\frac{\tilde
g}{V}\left[\sum_{{\bf k}\neq 0}\hat{a}_{\bf k}^\dag\hat{a}_{\bf
k}\right]^2.\label{Heff}
\end{eqnarray}
\end{widetext}
In the effective Hamiltonian (\ref{Heff}), the only processes
considered are the annihilation of a pair $\{{\bf k},-{\bf k}\}$
into the condensate through the two-body interaction, the scattering
of a single particle ${\bf k}$ into the condensate by disorder, and
the corresponding inverse processes. When the condensate becomes
depleted, the last term in Hamiltonian ({\ref{Heff}}) is important
\cite{Huang,Pitaevskii} and can be treated by making the replacement
$\frac{1}{V}\left[\sum_{{\bf k}\neq 0}\hat{a}_{\bf
k}^\dag\hat{a}_{\bf k}\right]^2\rightarrow n'\sum_{{\bf k}\neq
0}\hat{a}_{\bf k}^\dag\hat{a}_{\bf k}, $ where $n^{\prime}$ is a
parameter to be determined later. We emphasize here that the
introduction of $n'$ in Eq. (\ref{Heff}) is equivalent to the
introduction of an additional Lagrange multiplier as shown in the
Refs. \cite{Yukalov,Yukalovb,Yukalovd}. The effective Hamiltonian
(\ref{Heff}) is then diagonalized by the Bogoliubov transformation
\begin{eqnarray}
\hat{a}_{\bf k} &=&u_{\bf k}\hat{c} _{\bf k}-\upsilon _{\bf
k}\hat{c} _{-\bf k}^{\dag }-\frac{\sqrt{N_0}(u_{\bf k}-\upsilon
_{\bf
k})^2}{E_{\bf k}}V_{k},\nonumber\\
\hat{a}^{\dagger}_{\bf k} &=&u_{\bf k}\hat{c}^{\dagger} _{\bf
k}-\upsilon _{\bf k}\hat{c}_{-\bf k}-\frac{\sqrt{N_0}(u_{\bf
k}-\upsilon _{\bf k})^2}{E_{\bf k}}V_{-k},
\end{eqnarray}
with $\upsilon _{\bf k}^2=u_{\bf
k}^2-1=\frac{1}{2}\left(\frac{\varepsilon_k^0+n_0\tilde{g}}{E_{\bf
k}}-1\right)$, $E_{\bf
k}=\sqrt{(\varepsilon_k^0+\Delta)^2+2n_0\tilde{g}(\varepsilon_k^0+\Delta)}$,
and $\Delta=\tilde{g}\left(n_0+n^{\prime}\right)-\mu$. Consequently,
the diagonalized Hamiltonian reads:
\begin{equation}\label{Hdia}
H_{eff}-\mu N=-V\mu n_0+E_g+\sum_{{\bf k}\neq 0}E_{\bf
k}\hat{c}_{\bf k}^\dag\hat{c}_{\bf k},
\end{equation}
with the ground state energy $E_g$ reading
\begin{eqnarray}\label{Eg0}
E_g&=&\frac{1}{2}V\tilde{g}n_0^2-\frac{1}{2}\sum_{k\neq
0}(\varepsilon_k^0+n_0\tilde{g}-E_k)\nonumber\\
&+&N\left[n_{imp}\tilde{g}_{imp}-\frac{n_{imp}g_{imp}^2}{V}\sum_{k\neq
0}\frac{\varepsilon_k^0}{E_k^2}\right].
\end{eqnarray}
In conformity with general theorem \cite{Hugenholtz}, here we set
$\Delta=\tilde{g}(n_0+n^{\prime})-\mu=0$ to ensure a gapless
quasiparticle spectrum \cite{Yukalovb}. This condition, together
with $n=n_0+1/V\sum_{{\bf k}\neq 0}\hat{a}^{\dagger}_{{\bf
k}}\hat{a}_{{\bf k}}$, determines $n^{\prime}$ and $n_0$ as a
function of $n$.

\section{Quantum fluctuations and dimensional crossover}

\subsection{Beyond mean field correction to the ground state energy}

By replacing the sum in Eq. ({\ref{Eg0}}) with integrals in the
continuum limit, we obtain
\begin{eqnarray}
\frac{E_g}{V}&=&\frac{1}{2}\tilde{g}n_0^{2}-\frac{1}{ 4\pi
}\frac{\tilde{g}n_0}{d^{2}}\sqrt{2m\tilde{g}n_0}f\left(
\frac{2t}{\tilde{g}n_0}\right)\nonumber\\
\!\!&\!+\!&\!\!\left[ \frac{1}{2}\tilde{g}n_0
^{2}\left(\kappa\frac{\tilde{b}
}{\tilde{a}}\right)\!-\!\frac{\tilde{R}\tilde{g}n_0}{4\pi \hbar
d^{2}}\sqrt{mn_0\tilde{g}}Q\!\!\left(\frac{2t}{\tilde{g}n_0}\right)\right],
\label{Egav}
\end{eqnarray}
with $\kappa=n_{imp}/n_0$ and
$\tilde{R}=\kappa\tilde{b}^2/\tilde{a}^2$. The last two terms in Eq.
(\ref{Egav}) are the beyond mean-field correction to the ground
state energy due to disorder. The functions $f(x)$ and $Q(x)$ in Eq.
(\ref{Egav}) are respectively defined as
\begin{equation}
f(x)=\frac{\pi }{2\sqrt{x}}\int_{-\pi }^{\pi
}\frac{d^{2}\mathbf{{\bf k}}}{(2\pi
)^{2}}\frac{_{2}F_{1}(\frac{1}{2},\frac{3}{2},3,-\frac{2}{x{\gamma ({\bf k})}})}{\sqrt{%
\gamma({\bf k})}},\label{f}
\end{equation}
and
\begin{equation}
Q(x) =\frac{\pi}{2} \int_{-\pi}^{\pi}\frac{dk_{x}}{2\pi}\frac{
_{2}F_{1}(\frac{1}{2},\frac{1}{2},1,\frac{x^{2}}{\left( 1+x+x\sin
^{2}(k_{x}/2)\right) ^{2}})}{\sqrt{x\sin
^{2}\left(k_{x}/2\right)+1}}.\label{J}
\end{equation}
In both Eqs. (\ref{f}) and (\ref{J}), the variable $x$ stands for
$2t/\tilde{g}n_0$ that appears in Eq. (\ref{Egav}). In Eq.
(\ref{f}), $\gamma({\bf k})=2-\cos(k_x)-\cos(k_y)$. The function
$_2F_1(a,b,c,d)$ in Eqs. (\ref{f}) and (\ref{J}) is the
hypergeometric function \cite{Zwillinger} and the integration over
the transverse quasimomenta is restricted to the first Brillouin
zone, i.e. $|k_x|,|k_y|\leq \pi$ in Eq. (\ref{f}) and $|k_x|\leq\pi$
in Eq. (\ref{J}).

Eq. (\ref{Egav}) describes the ground state energy of a BEC in a
combined presence of tight 2D periodic potential and weak disorder.
Our model is characterized by three parameters: (i) the BEC
parameter $n\tilde{a}^3$, (ii) the concentration of disorder
$\kappa=n_{imp}/n_0$, and (iii) the ratio of effective scattering
amplitudes $\tilde{b}/\tilde{a}$. The first one reflects the
strength of interatomic interaction in the presence of periodic
potential, whereas the other two are important parameters that
describe the effect of disorder trapped in optical lattice.

\subsection{Lattice-induced dimensional crossover in the presence of weak disorder}

We now proceed to consider the asymptotic behavior of the ground
state in the limit of $x\rightarrow 0$ and $x \gg 1$, respectively.
In the limit of $x\gg 1$, corresponding to $8t\gg\mu$, the system
retains an anisotropic 3D behavior; whereas for $x\rightarrow 0$,
corresponding to $8t\ll\mu$, the system undergoes a dimensional
crossover to a 1D regime.

In the 1D regime $x\rightarrow 0$, the $f(x)$ saturates to the value
$4\sqrt{2}/3$ \cite{Orso}. In this limit, we can neglect the Bloch
dispersion and Eq. (\ref{Hdia}) approaches asymptotically to the
ground state energy of a 1D Bose gas in the presence of weak
disorder
\begin{eqnarray}
\frac{E_g}{L}&=&\frac{1}{2}g_{1D}n^2_{1D}-\frac{2}{3\pi}\sqrt{m}\left(n_{1D}g_{1D}\right)^{3/2}\nonumber\\
&+&\Big[\frac{1}{2}g_{1D}n^2_{1D}\left(\kappa\frac{\tilde{b}
}{\tilde{a}}\right)\nonumber\\
&&-\frac{\tilde{R}\sqrt{m}}{4\pi\hbar
}(n_{1D}g_{1D})^{3/2}Q\left(\frac{2t}{g_{1D}n_{1D}}\right)\Big],\label{E1D}
\end{eqnarray}
with linear density $n_{1D}=n_0d^2$ and coupling constant
$g_{1D}=\tilde{g}/d^2$. Here $L$ is the length of the tube. In the
case of vanishing disorder ($\kappa=0$), Eq. (\ref{E1D}) is in
agreement with the exact Lieb-Linger solution of the 1D model
expanded in the weak coupling regime $mg_{1D}/\hbar n_{1D}\ll 1$
\cite{Lieb}. We argue, therefore, that Eq. (\ref{E1D}) with
$\kappa\neq 0$ is the generalized Lieb-Linger solution of the 1D
model expanded in the weak coupling regime in the presence of weak
disorder.

In the opposite 3D regime $x\gg 1$, the functions (\ref{f}) and
(\ref{J}) approach respectively the asymptotic law $f(x)\simeq
1.43/\sqrt{x}-16\sqrt{2}/(15\pi x)$ and $Q(x)\simeq -1/\sqrt{\pi}x$.
Hence in the limit $x\gg 1$, Eq. (\ref{Egav}) takes the asymptotic
form
\begin{eqnarray}
\frac{E_g}{V}=\frac{2\pi\hbar^2}{m}n_0^2\tilde
a\Big[\Big(1\!&+&\!\!\kappa
\frac{\tilde{b}}{\tilde{a}}\Big)+\frac{\tilde a}{\tilde
a_{cr}}+\frac{128}{15}\left(\frac{n_0\tilde{a}^3}{\pi}\right)^{1/2}\!\!\frac{m^{*}}{m}\nonumber
\\
&+&4\sqrt{\pi}\left(\frac{n_0\tilde{a}^3}{\pi}\right)^{1/2}\frac{m^{*}}{m}\tilde{R}\Big],\label{E3D}
\end{eqnarray}
with $m^{*}=\hbar^2/2td^2$ being the effective mass \cite{Kramer}
associated with the band and $\tilde a_{cr}=-0.24d\sqrt{m/m^{*}}$
being a further renormalization of the scattering length due to the
optical lattice. \cite{Orso}. The first two terms in Eq. (\ref{E3D})
are the mean-field contribution; whereas the remaining terms are
generalized LHY correction in the presence of both optical lattice
and disorder. In particular, the presence of disorder gives rise to
the last term in Eq. (\ref{E3D}). Compared to the free case, the
periodic potentials affect the ground state energy in two ways.
First, the periodic potential increases the local atomic density,
thereby enhancing interatomic interactions characterized by the
effective scattering length $\tilde a=C^2a$. Second, the lattice
modifies the dispersion relation, which is captured by the effective
mass $m^{*}$. For vanishing disorder $\kappa=0$, Eq. (\ref{E3D}) is
reduced to the corresponding result in Ref. \cite{Orso}; whereas for
vanishing optical lattice $s=0$, our result recovers exactly the
corresponding one in Ref. \cite{Astra}.

\subsection{Quantum depletion}

The average number of particles with nonzero momentum represents a
depletion of the condensate which can be calculated by
$N-N_0=\sum_{{\bf k}\neq0}\left(\epsilon^{0}_{{\bf k}}+\tilde g n_0-
E_{{\bf k}}\right)/2E_{{\bf k}}$. In the continuum limit, the
quantum depletion of a BEC trapped in a 2D optical lattice in the
presence of disorder is calculated as
\begin{equation}
\!\!\frac{N-N_0}{N}\!=\!\frac{1}{4\pi d^{2}}\sqrt{\frac{2m
\tilde{g}}{n_0}}h\left( \frac{2t}{\tilde{g}n_0}\right)
+\!\frac{\tilde{R}}{16\pi \hbar
d^{2}}\sqrt{\frac{m\tilde{g}}{n_0}}K\!\!\left(
\frac{2t}{\tilde{g}n_0}\right), \label{DNav}
\end{equation}
where $h(x)$ and $K(x)$ are functions of the variable
$x=2t/\tilde{g}n_0$ that are respectively defined as
\begin{equation}
h(x)\!=\!\int_{-\pi }^{\pi }\frac{d^{2}{\bf k}}{(2\pi )^{2}}\int_{x{%
\gamma}({\bf k})}^{+\infty }\frac{dt}{\sqrt{t-x{\gamma}({\bf k})}}%
\left[ \frac{t+1}{\sqrt{t^{2}+2t}}-1\right],\label{h}
\end{equation}
and
\begin{widetext}
\begin{equation}
K(x)=\frac{\pi}{2}\int_{-\pi}^{\pi}\frac{dk}{2\pi}\frac{
_{2}F_{1}(\frac{1}{2},\frac{1}{2},1,\frac{x^{2}}{\left[ 1+x+x\sin
^{2}\left(k/2\right)\right] ^{2}})}{\left[1+x+x\sin
^{2}(k/2)\right]\sqrt{x\sin ^{2}(k/2)+1}}.\label{K}
\end{equation}
\end{widetext}
The first term in Eq. (\ref{DNav}) arises from the two-body
interactions of bosonic atoms trapped in the optical lattice. The
second term, on the other hand, represents the effect of disorder on
the quantum depletion.

In the 1D regime, $h(x)\simeq -\ln(2.7x)/\sqrt{2}$ so the first term
in Eq. (\ref{DNav}) diverges. It implies that in the absence of
tunneling no real Bose-Einstein condensation exists which agrees
with the general theorems in one dimension \cite{Pitaevskii}. In the
opposite 3D regime, the functions $h(x)$ and $K(x)$ respectively
decay as $4/(3\pi\sqrt{2}x)$ and $2/(\pi x)$. Accordingly, the
quantum depletion (\ref{DNav}) takes the asymptotic form in the 3D
regime:
\begin{equation}
\frac{N-N_0}{N}=\left(\frac{8}{3}+\frac{\tilde{R}\pi}{2}\right)\left(\frac{m^{*}}{m}\right)\left(\frac{n_0\tilde
a^3}{\pi}\right)^{1/2}. \label{DN3D}
\end{equation}
Eq. (\ref{DN3D}) generalizes the standard 3D result in the presence
of weak disorder in free space to that in the presence of an optical
lattice.

\section{Reduction of superfluid density due to combined effects of disorder and optical lattice}
\subsection{Longitudinal and transverse response functions}

In the second part of this paper, we apply the linear response
theory to investigate the effects of disorder on the superfluid
density of a BEC trapped in a 2D optical lattice. The general
definition of the superfluid density is proposed by Hohenberg and
Martin \cite{Hohenberg}. We emphasize that, unlike the condensate
density, superfluidity is a kinetic property of a system and the
superfluid density is not an equilibrium quantity but a transport
coefficient, and should be determined by the response of momentum
density to an externally imposed velocity field
\cite{Hohenberg,Forster}. According to Kubo's formula \cite{Liquid},
the average momentum density $\langle g({\bf r},t)\rangle$ induced
by the external perturbation velocity field $\delta v({\bf r^{'}})$
is given by
\begin{equation}
\left\langle g_i({\bf
r},t)\right\rangle=\int^{t}_{-\infty}dt^{'}\int d{\bf r'}R_{ij}({\bf
r}t,{\bf r}^{'}t^{'})\delta v_{j}({\bf r}^{'})e^{\epsilon t^{'}},
\end{equation}
where
\begin{equation}
 g({\bf r},t)=\frac{\hbar}{2
i}\left[\hat{\Psi}^{\dagger}({\bf r})\nabla \hat{\Psi}({\bf
r})-\hat{\Psi}({\bf r})\nabla \hat{\Psi}^{\dagger}({\bf r})\right],
\end{equation}
is the momentum density operator, and
\begin{eqnarray}
\!\!\!\!\!\!R_{ij}({\bf r}t,{\bf
r}^{'}t^{'})\!\!&=&\!\!\left\langle\left[g_i({\bf
r},t),g_j({\bf r}^{'},t^{'})\right]\right\rangle\nonumber \\
\!\!\!&=&\!\!\!\!\int\! \frac{d{\bf
q}}{(2\pi)^3}\frac{d\omega}{2\pi}e^{i\left[{\bf q}\cdot({\bf
r}\!-\!{\bf r}^{'})\!-\!\omega(t\!-\!t^{'})\right]}R_{ij}({\bf
q},\omega),
\end{eqnarray}
is the momentum density correlation function, averaged with the
ground state wavefunction for a BEC at rest without perturbation at
zero temperature. The static current-current response function is
given by \cite{Baym}
\begin{eqnarray}
\chi_{ij}({\bf
q})&=&\frac{1}{m}\int^{+\infty}_{-\infty}\frac{d\omega}{2\pi}\frac{R_{ij}({\bf
q},\omega)}{\omega-i\epsilon}\nonumber\\
&=&-2\sum_n\frac{\left\langle0|J_{\bf
q}^i|n\right\rangle\left\langle n| J_{-\bf
q}^j|0\right\rangle}{\omega_{n0}},
\end{eqnarray}
where $\omega_{n0}=(\varepsilon_n-\varepsilon_0)/\hbar$ and
$|0\rangle$ is the ground state and the sum is performed over a
complete set of excited states $|n\rangle$ with the energy
$\hbar\omega_n$. Here the current operator $J_{\bf q}^i$ is defined
by
\begin{eqnarray}
J_{\bf q}^{x(y)}&=&\frac{2td}{\hbar}\sum_{\bf
k}\sin\left(k_{x(y)}+\frac{q_{x(y)}}{2}\right)\hat{a}^{\dagger}_{\bf
k}a_{\bf k+q},\label{jx}\\
J_{\bf q}^{z}&=&\frac{\hbar}{2m}\sum_{\bf
k}\left(k_{z}+\frac{q_{z}}{2}\right)\hat{a}^{\dagger}_{\bf k}a_{\bf
k+q}.\label{jz}
\end{eqnarray}

In general, the superfluid density of a BEC distinguishes between
the low-frequency, long-wavelength longitudinal and transverse
responses of the system \cite{Huang,Pitaevskii,Forster,Liquid,Baym}
\begin{equation}\label{lt}
\chi_{ij}\left({\bf q }\right)=\frac{q_iq_j}{q^2}\chi_{L}\left({\bf
q
}\right)+\left(\delta_{ij}-\frac{q_iq_j}{q^2}\right)\chi_{T}\left({\bf
q }\right).
\end{equation}
The first term in Eq. ({\ref{lt}}) is the longitudinal component; it
is parallel to ${\bf q}$ in both indices, i.e.
$\sum_{i}q_i(q_iq_j/q^2)=q_j$. The second term, the transverse
component is perpendicular to ${\bf q}$ in both indices, i.e.
$\sum_{i}q_i(\delta_{ij}-q_iq_j/q^2)=0$.

A longitudinal probe corresponds to boosting the system and the
entire fluid responds. On the other hand, a low-frequency transverse
probe corresponds to a slow rotation of the system, which only
couples to the normal fluid while leaving the superfluid untouched.
Hence the normal fluid density is defined in terms of the static
transverse current-current correlation function as
\cite{Huang,Pitaevskii,Forster,Liquid,Baym}
\begin{equation}
\rho_n =\lim_{q\rightarrow 0}\chi_{T}({\bf q}).\label{Normal}
\end{equation}
The superfluid density is thereby defined as the difference between
the total density $\rho$ and the normal fluid density
\begin{equation}
\rho_s =\rho-\lim_{q\rightarrow 0}\chi_{T}({\bf
q}).\label{Superfluid}
\end{equation}
There is an important point to stress here. By using the particle
conservation, one can obtain $\rho=\lim_{q\rightarrow
0}\chi_{L}({\bf q})$ \cite{Huang,Pitaevskii,Forster,Liquid,Baym}.
This result is fixed by the model-independent $f$-sum rule
\cite{Pitaevskii,Forster,Liquid,Baym}. One is therefore easily led
to directly write $\rho_s =\lim_{q\rightarrow 0}\left[\chi_{L}({\bf
q})-\chi_{T}({\bf q})\right]$. This is not valid, however, for the
calculation using the Bogoliubov approximation which violates the
particle conservation by replacing $\hat{a}_0$ with a {\it c}-number
$\sqrt{N_0}$ \cite{Huang}. In particular, computations involving the
time evolution of zero-momentum particle states would be false. The
non-zero momentum states, on the other hand, are not affected. Thus
for our present calculation of the superfluid density within the
Bogoliubov' theoretical framework, it is the calculation of Eq.
(\ref{Normal}) that can be trusted, not $\lim_{q\rightarrow
0}\chi_{L}({\bf q})$. Accordingly, the superfluid density can only
be determined through Eq. ({\ref{Superfluid}}) in the present
context.

\subsection{Superfluid density}
The transverse response of a BEC is only due to the normal fluid,
since the superfluid component can only participate in irrotational
flow. For a BEC trapped in a 2D optical lattice, the rotational
symmetry is broken that gives rise to an anisotropic system.
Consequently, the response in the unconfined z-direction is
different from that in the confined x-y plane. Nevertheless, the
system is still isotropic in the x-y plane. Hence it's more
convenient to consider a slow rotation with respect to the z-axis.
We therefore take the transverse response function along the
z-direction, given by
\begin{eqnarray}\label{chizz}
\chi _{zz}(0)&=& 2\sum_{{\bf k}}\frac{\hbar^2k_z^2}{m}\frac{\varepsilon_{\bf k}^0}{E_{{\bf k}}^4}\langle|V_{{\bf k}}|^2\rangle\nonumber\\
&=&\frac{\tilde{R}}{4\hbar
d^2}\sqrt{2m\tilde{g}n_0}I(\frac{2t}{\tilde{g}n_0}),
\end{eqnarray}
with $I(x)$ being a function of a variable $x=2t/\tilde{g}n_0$ in
the form
\begin{equation}
\!\!\!\!\!I(x)\!\!=\!\!\int_{-\pi }^{\pi }\frac{d^2{\bf
k}}{(2\pi)^2}\frac{1}{\sqrt{x\gamma({\bf k}) +2}}\frac{1}{\left(
\sqrt{x}+\sqrt{x\gamma({\bf k})+2}\right) ^{2}}.
\end{equation}
Two properties of the normal fluid density can immediately be stated
based on Eq. (\ref{chizz}): (i) the normal fluid component induced
by disorder is not fixed by the long-wavelength properties of the
excited quasi-particle. In order to assure the convergence of the
integral in Eq. (\ref{chizz}), the behavior of the elementary
excitation spectrum at high momenta is important; (ii) Eq.
(\ref{chizz}) can be interpreted as the second-order term in the
perturbation expansion of the normal fluid density in terms of weak
disorder $V_{{\bf k}}$. These two properties should be general and
do not depend on the concrete form of disorder \cite{Huang}.

The superfluid density is thereby found to be $\rho_s=\rho-\rho_n$
with $\rho_n=\chi _{zz}(0) $. In the region $x\gg 1$, $I(x)$
asymptotically approaches the function
 $\sqrt{2}/(6\pi x)$. Eq. (\ref{chizz}) therefore takes the
 asymptotic form
 \begin{equation}\label{Normal3D}
\chi _{zz}(0)= \frac{m^{\ast }}{m}\frac{2\sqrt{\pi}}{3}
\tilde{R}n_0\left( n_0\tilde{a}^{3}\right) ^{1/2}.
 \end{equation}
Eq. (\ref{Normal3D}) generalizes the normal fluid density induced by
disorder in the free space to that in the presence of optical
lattice. Accordingly, we conclude that in the asymptotic anisotropic
3D regime, the disorder still generates an amount of normal fluid
that is equal to $4/3$ of the condensate depletion \cite{Huang},
despite the presence of optical lattice.

To conclude this section, we note that in using Eq. ({\ref
{Superfluid}}) to determine the superfluid density, one has to
specify detailed probing approach and the limiting procedures
thereof to distinguish transverse and longitudinal responses
\cite{Huang, Pitaevskii,Baym}. In rotationally invariant and uniform
systems, definition ({\ref {Superfluid}}) has been a standard
expression to calculate the superfluid density
{\cite{Huang,Pitaevskii}}. It would be less advantageous, however,
for nonuniform systems. There are other definitions for superfluid
density {\cite{Yukalov,GSuper1,GSuper2}}. Despite the formal
difference, these various definitions are fundamentally grounded in
analyzing the linear response of a fluid to an external velocity
boost. The underlying key quantity to determine superfluid
components, therefore, is the momentum density correlation function.
In isotropic and uniform systems, these definitions all lead to the
same superfluid fraction within the linear response approximation.
To determine the superfluid density in nonuniform systems, Refs.
{\cite{Yukalov,GSuper1,GSuper2}} gave a more general expression. In
the present context, both the definition ({\ref{Superfluid}}) and
the definition in Refs. {\cite{Yukalov,GSuper1,GSuper2}} are
appropriate. This is because the presence of optical lattice has
been accounted for through renormalizing the scattering length
$\tilde{g}$ and the kinetic energy term $\varepsilon^0_{{\bf k}} $
\cite{Kramer}. As a result, despite the presence of lattice, one can
study the quantum behaviors of a BEC as if the space is homogeneous
\cite{Kramer,Kramera}. In particular, in the limit $8t\gg\mu$ the
effect of lattice is explicitly captured by the effective mass
$m^{*}$ and the effective interaction $\tilde{g}$. Consequently in
the asymptotic 3D region, a BEC trapped in an optical lattice can be
effectively regarded as a uniform fluid composed of bosons with an
effective mass $m^{*}$ and an effective coupling constant
$\tilde{g}$ \cite{Orso,Kramer}. Thus within our present context, Eq.
({\ref{Superfluid}}) still gives a good definition for the
superfluid density particularly in the asymptotic 3D region, despite
the presence of lattice.

\section{Possible Experimental Scenarios and Conclusion}

In this work, the physics of our model is captured by the interplay
among three quantities: the strength of an optical lattice $s$, the
interaction between bosonic atoms $\tilde{g}n_0$, and the strength
of disorder $\tilde{R}$. All these quantities are experimentally
controllable using state-of-the-art technologies. The interatomic
interaction can be controlled in a very versatile manner via the
technology of Feshbach resonances. In the typical experiments to
date, the values of ratio $\tilde{g}n_0/E_R$ range from $0.02$ to
$1$ \cite{Morschrev,Blochrev}. The depth of an optical lattice $s$
can be changed from $0E_R$ to $32E_R$ almost at will \cite{Greiner}.
Disorder may be created in a repeatable way by introducing
impurities in the sample \cite{Ospelkaus}, or using laser speckles
and multi-chromatic lattices \cite{Damski,Roth,Lye,White}.
Therefore, the phenomena discussed in this paper should be
observable within the current experimental capability.

We emphasize here that our investigation of a BEC in the presence of
both optical lattice and weak disorder has been done within the
Bogoliubov theory. Further improvement of the theoretical framework
is to permit the treatment of the system properties for whole range
of interatomic interaction strength, from zero to infinity, as well
as arbitrarily strong disorder \cite{Yukalovd}.

In summary, a BEC trapped in a 2D optical lattice in the presence of
disorder is analytically investigated within the framework of the
Bogoliubov theory. We focus on analyzing the combined effects of
disorder and optical lattice on the quantum fluctuations and
superfluid density of the BEC system. Accordingly, the analytical
expressions of the ground state energy and the quantum depletion of
the system are obtained. Our results show that the 2D lattice
induces a characteristic 3D to 1D crossover in the behavior of
quantum fluctuations. Furthermore, the effects of disorder on this
dimensional crossover are discussed in detail. Especially, we argue
that our results in the asymptotic 1D regime is a generalized
Lieb-Linger solution of the 1D model expanded in the weak coupling
regime in the presence of weak disorder. In addition, the normal
fluid density of the condensate induced by the disorder is obtained
by calculating the transverse response function. Our results show
that in the 3D regime, the quantum depletion due to the combined
effects of disorder and lattice will asymptotically approach $3/4$
of the corresponding reduction of the superfluid density. Finally,
the conditions for possible experimental realization of our scenario
are discussed.
\newline

\textbf{Acknowledgements} We thank Biao Wu for helpful discussions.
This work is supported by the Hongkong Research Council (RGC) and
the Hong Kong Baptist University Faculty Research Grant (FRG). ZXL
is supported by the IMR SYNL-TS Ke Research Fellowship.


\begin{thebibliography}{99}
\bibitem{Morschrev}{O. Morsch  and M. Oberthaler, Rev. Mod. Phys. \textbf{78},
179 (2006).}
\bibitem{Blochrev} {I. Bloch, J. Dalibard and W. Zwerger,Rev. Mod.
Phys. \textbf{80}, 885 (2008).}
\bibitem{Yukalov}{V. I. Yukalov, Laser Phys. \textbf{19}, 1 (2009).}
\bibitem{Greiner}{M. Greiner, O. Mandel, T. Esslinger, T. W. Hansch, and J. Bloch, Nature
(London) {\bf 415}, 39, (2002).}
\bibitem{Tiesinga}{E. Tiesinga, B.J. Verhaar, H.T.C. Stoof, Phys. Rev. A. {\bf 47}, 4114
(1993); S. Inouye {\it et al.}, Nature (London) {\bf 392}, 151
(1998).}
\bibitem{Orso}{G. Orso, C. Menotti, and S. Stringari, Phys. Rev. Lett. {\bf 97}, 190408 (2006).}
\bibitem{Yukalovd}{V. I. Yukalov and E. P. Yukalova, Phys. Rev. A {\bf 74}, 063623 (2006); V. I. Yukalov, E. P. Yukalova, K. V. Krutitsky, and R. Graham,
{\it ibid}. {\bf 76}, 053623 (2007).}
\bibitem{Kuhn} {R. C. Kuhn, C. Miniatura, D. Delande, O.
Sigwarth, and C. A. M\"uller, Phys. Rev. Lett. {\bf 95}, 250403
(2005).}
\bibitem{Gimperlein}{H. Gimperlein, S. Wessel, J. Schmiedmayer, and L. Santos, Phys. Rev. Lett. {\bf 95}, 170401 (2005).}
\bibitem{Sanchez}{L. Sanchez-Palencia, D. Cl\'ement, P. Lugan, P. Bouyer, G.V. Shlyapnikov, and A. Aspect, Phys. Rev. Lett. {\bf 98}, 210401 (2007).}
\bibitem{Delande}{D. Delande and J. Zakrzewski, Phys. Rev. Lett. {\bf 102}, 085301 (2009).}
\bibitem{Pilati}{S. Pilati, S. Giorgini, and N. Prokofev, Phys. Rev. Lett. {\bf 102}, 150402 (2009).}
\bibitem{Fontanesi}{L. Fontanesi, M. Wouters, and V. Savona,  Phys. Rev. Lett. {\bf 103}, 030403 (2009).}
\bibitem{Dang}{L. Dang, M. Boninsegni, and L. Pollet, Phys. Rev. B {\bf 79}, 214529 (2009).}
\bibitem{Falco}{G. M. Falco, T. Nattermann, and V. L. Pokrovsky, Europhys. Lett. {\bf 85}, 30002 (2009).}
\bibitem{Wang}{D. W. Wang, M. D. Lukin, and E. Demler, Phys. Rev. Letts. {\bf 92}, 076802 (2004).}
\bibitem{Lieb}{E. H. Lieb and W. Liniger, Phys. Rev. {\bf 130}, 1605 (1963); E. H. Lieb, {\it ibid.} {\bf 130}, 1616 (1963).}
\bibitem{Huang}{Kerson Huang and Hsin-Fei Meng, Phys. Rev. Lett. {\bf 69}, 644 (1992); S. Giorgini, L. Pitaevskii, and S. Stringari, Phys. Rev. B {\bf 49}, 12938 (1994);
A. V. Lopatin and V. M. Vinokur, Phys. Rev. Lett. {\bf 88}, 235503
(2002).}
\bibitem{Damski}{B. Damski, J. Zakrzewski, L. Santos, P. Zoller, and M. Lewenstein, Phys. Rev. Lett. {\bf 91}, 080403
(2003).}
\bibitem{Roth}{R. Roth and K. Burnett, Phys. Rev. A. {\bf 68}, 023604 (2003).}
\bibitem{Ospelkaus}{S. Ospelkaus, C. Ospelkaus, O. Wille, M. Succo, P. Ernst, K. Sengstock, and K. Bongs, Phys. Rev. Lett. {\bf 96}, 180403 (2006). }
\bibitem{Gavish}{U. Gavish and Y. Castin, Phys. Rev. Lett. {\bf 95}, 020401 (2005).}
\bibitem{Lye}{J. E. Lye, L. Fallani, M. Modugno, D. S. Wiersma, C. Fort, and M. Inguscio, Phys. Rev. Lett. {\bf 95}, 070401 (2005).}
\bibitem{White}{M. White, M. Pasienski, D. McKay, S. Q. Zhou, D. Ceperley, and B. DeMarco, Phys. Rev. Lett. {\bf 102}, 055301 (2009).}
\bibitem{Billy}{J. Billy {\it et al}., Nature (London) {\bf 453}, 891 (2008); G. Roati {\it et al}., Nature (London) {\bf 453}, 895 (2008). }
\bibitem{Astra}{G. E. Astrakharchik, J. Boronat, J. Casulleras, and S. Giorgini, Phys. Rev. A {\bf 66}, 023603 (2002).}
\bibitem{Wouters}{M. Wouters and G. Orso, Phys. Rev. A {\bf 73}, 012707 (2006).}
\bibitem{Kramer}{M. Kr\"amer, C. Menotti, L. Pitaevskii, and S. Stringari, Eur. Phys. J. D {\bf 27}, 247 (2003).}
\bibitem{Kramera}{M. Kr\"amer, L. Pitaevskii, and S. Stringari, Phys. Rev. Lett. {\bf 88}, 180404 (2002).}
\bibitem{Pitaevskii}{L. Pitaevskii and S. Stringari, {\it Bose-Einstein Condensate} (Clarendon Press, Oxford, 2003).}
\bibitem{Yukalovb}{V. I. Yukalov and H. Kleinert, Phys. Rev. A {\bf 73}, 063612 (2006).}
\bibitem{Hugenholtz}{N. M. Hugenholtz and D. Pines, Phys. Rev. {\bf 116}, 489 (1959).}
\bibitem{Zwillinger}{D. Zwillinger, {\it Standard Mathematical Tables and Formulae}, (CRC Press, Boston, 1996).}
\bibitem{Hohenberg}{P. C. Hohenberg and P. C. Martin, Ann. Physics (NY) {\bf 34}, 291 (1965).}
\bibitem{Forster}{D. Forster, {\it Hydrodynamic Fluctuations, Broken Symmetry, and Correclations Functions} (Benjamin, Reading, MA, 1975).}
\bibitem{Liquid}{D. Pines and P. Nozi\'eres, {\it The Theory of Quantum Liquids}, (Benjamin, New York, 1966), Vol. I;
P. Nozi\'eres and D. Pines, {\it The Theory of Quatum Liquids},
(Addison-Wesley, Reading, MA, 1990), Vol. II.}
\bibitem{Baym}{G. Baym, {\it Mathematical Methods in Solid State and Superfluid Theory},
edited by R. C. Clark and G. H. Derrick (Oliver and Boyd, Edinburgh,
1969). p. 151.}
\bibitem{GSuper1}{V. I. Yukalov, Laser Phys. Lett. {\bf 1}, 435 (2004).}
\bibitem{GSuper2}{V. I. Yukalov, Ann. Phys. (N.Y.) {\bf 323}, 461 (2008).}




\end{thebibliography}
\end{document}